\documentclass[aps,prb,twocolumn,superscriptaddress,showpacs,showkeys]{revtex4-1}
\usepackage{graphicx}
\usepackage[abs]{overpic}
\usepackage{upgreek}
\usepackage{pdfpages}
\begin{document}
\newcommand{\nv}{NV$^-$}
\newcommand{\ucb}{Department of Physics, University of California, Berkeley, CA 94720-7300, USA}
\newcommand{\sst}{${^1}E\rightarrow{^1}A_1$}
\newcommand{\anu}{Laser Physics Centre, RSPE, Australian National University, Canberra, ACT 0200, Australia}
\title{The Infrared Absorption Band and Vibronic Structure of the Nitrogen-Vacancy Center in Diamond}
\author{P. Kehayias}
\email[]{pkehayias@berkeley.edu}
\affiliation{\ucb}
\author{M.W. Doherty}
\affiliation{\anu}
\author{D. English}
\affiliation{\ucb}
\author{R. Fischer}
\affiliation{Department of Physics, Technion - Israel Institute of Technology, Haifa 32000, Israel}
\author{A. Jarmola}
\affiliation{\ucb}
\author{K. Jensen}
\affiliation{\ucb}
\author{N. Leefer}
\affiliation{\ucb}
\author{P. Hemmer}
\affiliation{Department of Electrical and Computer Engineering, Texas A\&M University, College Station, TX 77843, USA}
\author{N.B. Manson}
\affiliation{\anu}
\author{D. Budker}
\email[]{budker@berkeley.edu}
\affiliation{\ucb}
\affiliation{Nuclear Science Division, Lawrence Berkeley National Laboratory, Berkeley CA 94720, USA}
\date{\today}
\begin{abstract}
Negatively-charged nitrogen-vacancy (NV$^-$) color centers in diamond have generated much interest for use in quantum technology. Despite the progress made in developing their applications, many questions about the basic properties of NV$^-$ centers remain unresolved. Understanding these properties can validate theoretical models of NV$^-$, improve their use in applications, and support their development into competitive quantum devices. In particular, knowledge of the phonon modes of the $^1A_1$ electronic state is key for understanding the optical pumping process. Using pump-probe spectroscopy, we measured the phonon sideband of the ${^1}E\rightarrow{^1}A_1$ electronic transition in the NV$^-$ center. From this we calculated the ${^1}E\rightarrow{^1}A_1$ one-phonon absorption spectrum and found it to differ from that of the ${^3}E\rightarrow{^3}A_2$ transition, a result which is not anticipated by previous group-theoretical models of the NV$^-$ electronic states. We identified a high-energy 169 meV localized phonon mode of the $^1A_1$ level.
\end{abstract}

\maketitle
\section{Introduction}
\begin{figure}[h]
\begin{center}
\begin{tabular}{cc}
\begin{overpic}[width=0.20\textwidth]{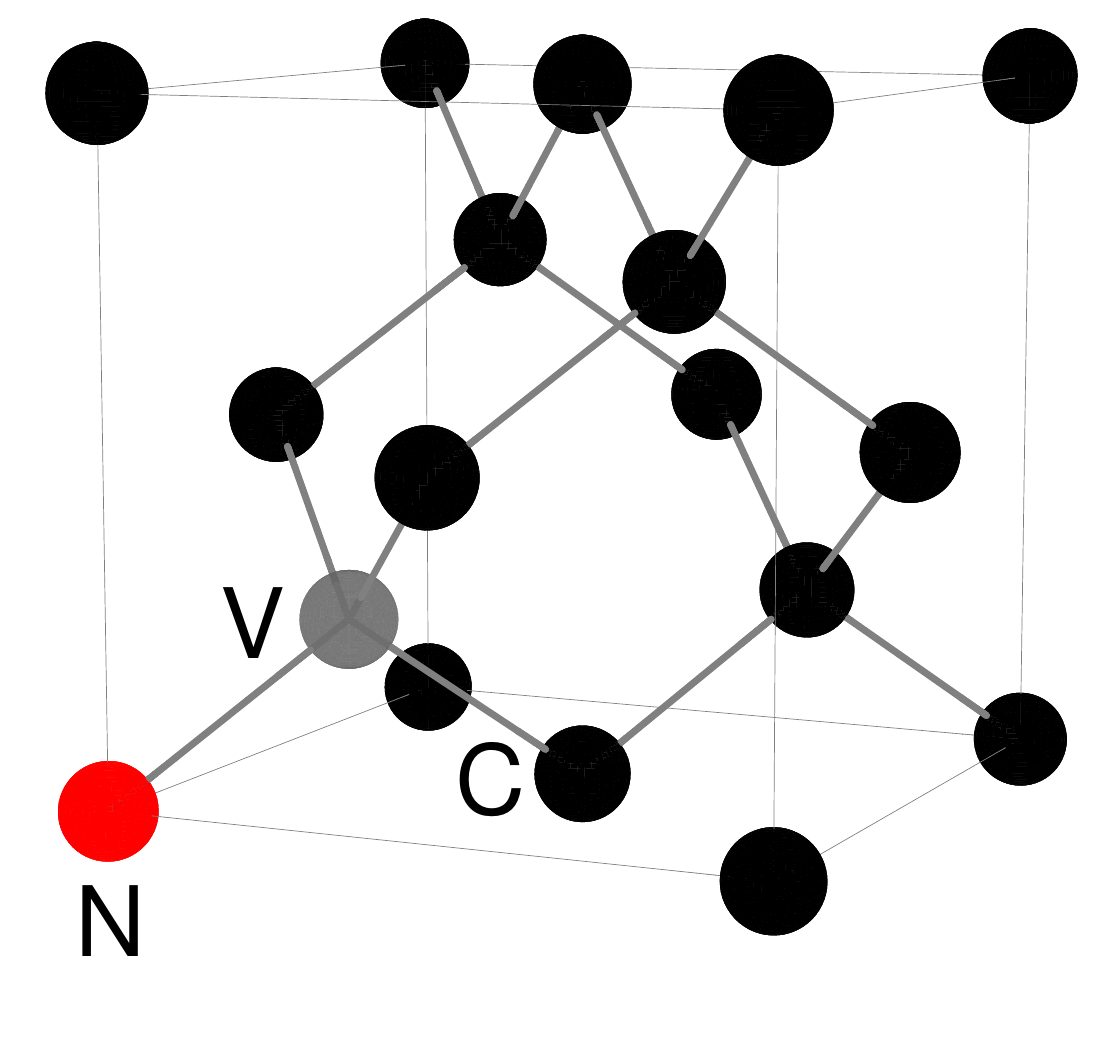}
\put(5,100){a}
\end{overpic}
&
\begin{overpic}[width=0.25\textwidth]{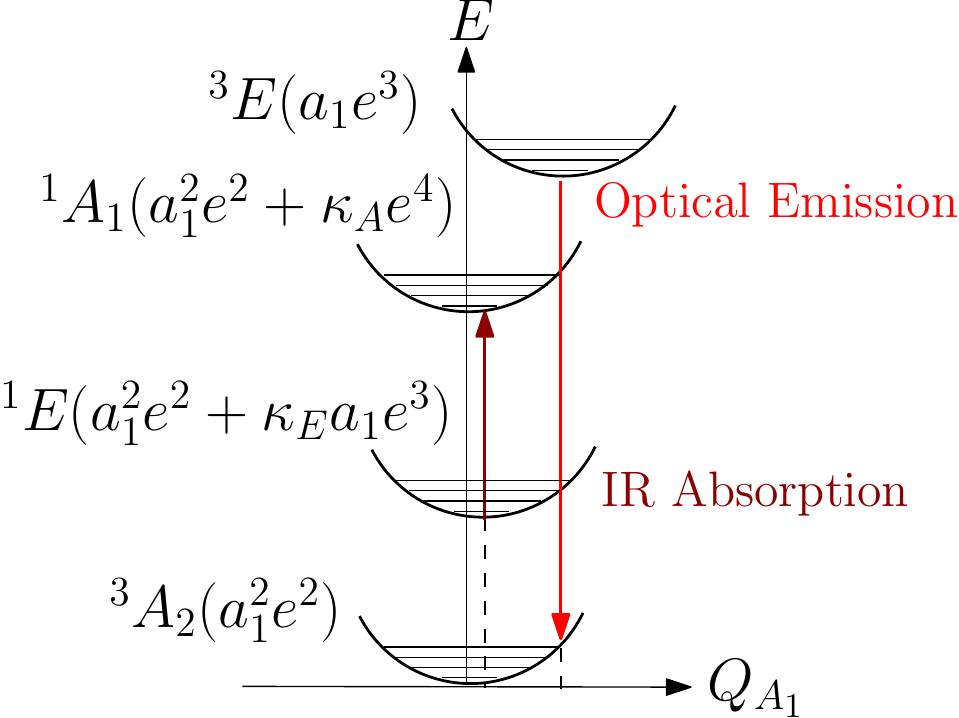}
\put(5,100){c}
\end{overpic}
\end{tabular}
\begin{overpic}[width=0.48\textwidth]{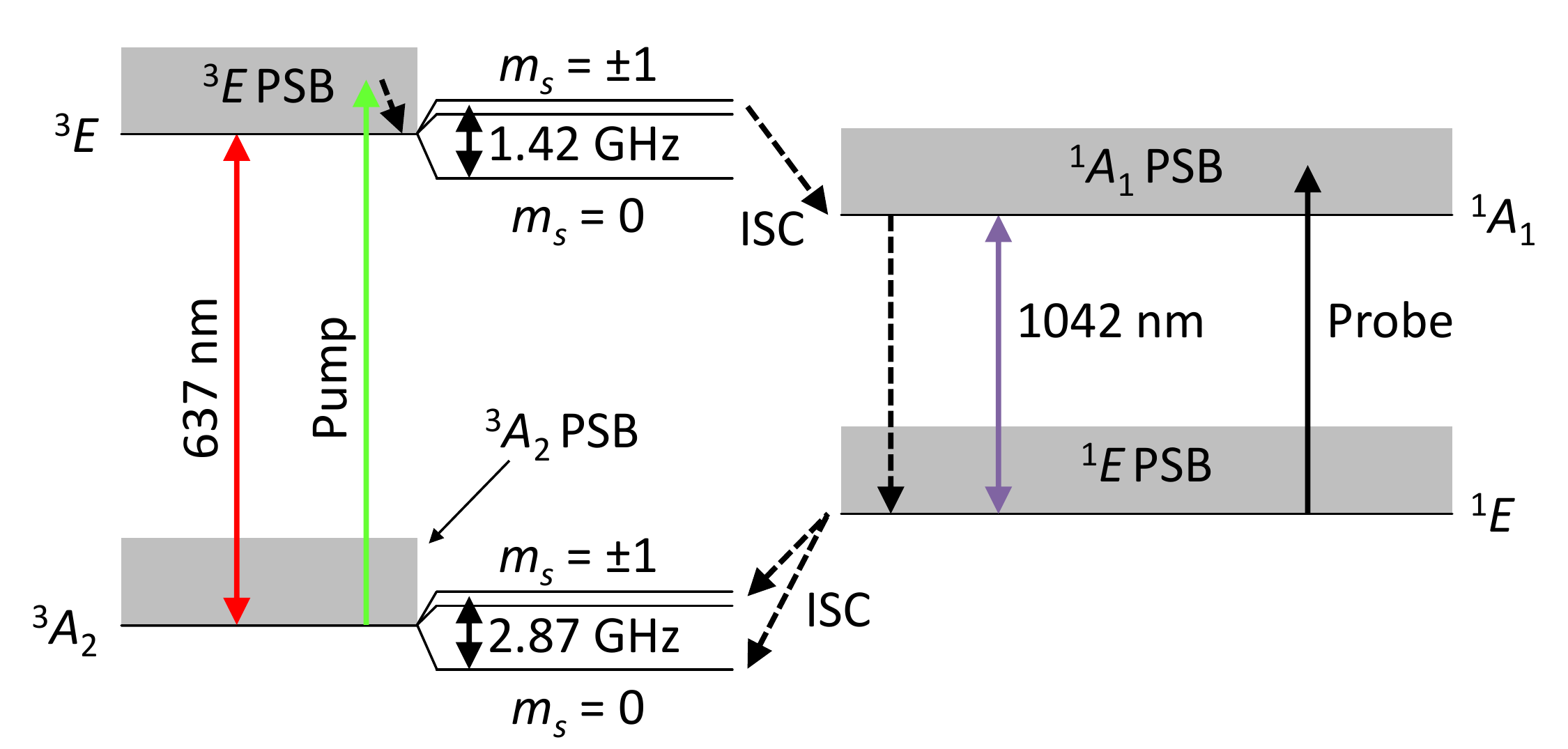}
\put(5,110){b}
\end{overpic}
\end{center}
\caption{\label{fig1}
(a) The diamond lattice structure, containing an NV center.
(b) The \nv\ energy-level diagram and our pump-probe spectroscopy scheme. The states are labeled by their $C_{3v}$ representations and electron spin multiplicities. Solid arrows are optical and microwave transitions, and dashed arrows are non-radiative transitions. The label ``ISC" indicates intersystem crossing, which occurs primarily for the $^3E$ $m_s = \pm 1$ states and is responsible for optical pumping.
(c) A configuration coordinate diagram for $A_1$ phonon modes showing the harmonic nuclear potential wells and phonon energy levels. The configuration for each electronic state is denoted in parentheses, and $Q_{A_1}$ is the normal nuclear coordinate. 
With no electronic Coulomb repulsion, the $^3A_2$, $^1E$, and $^1A_1$ levels are of the $a_1^2e^2$ configuration and the $^3E$ level is of the $a_1e^3$ configuration. With Coulomb repulsion included to first order, the $^1E$ and $^1A_1$ levels couple with the $^1E^\prime$ (configuration $a_1e^3$) and $^1A_1^\prime$ (configuration $e^4$) levels, respectively. This coupling is denoted by the parameters $\kappa_E$ and $\kappa_A$.
}
\end{figure}

The nitrogen-vacancy (NV) center in diamond (Fig.~\ref{fig1}a) is a color center consisting of a substitutional nitrogen atom in the diamond crystal lattice adjacent to a missing carbon atom (a vacancy). NV centers have $C_{3v}$ point-group symmetry and have discrete electronic energy states between the diamond valence and conduction bands. The negatively-charged \nv\ center can be optically spin-polarized and read out, and it has a long ground-state transverse spin relaxation time at room temperature.\cite{longT2, nir_CPMG} These properties make \nv\ centers useful in a variety of applications including electric and magnetic field sensing,\cite{dima_magnetometry_review, singleNV_mag, nano_img_mag, eFieldSensing} rotation sensing, \cite{gyros1a, gyros2, gyros3} quantum computing, \cite{jelezko_QIP, lukin_qubit} quantum cryptography, \cite{crypt1, crypt2} and sub-diffraction-limited imaging. \cite{STED1, STED2, STED3} Despite the progress made on developing these applications, the complete \nv\ energy level structure and vibronic structure are unknown.

Figure \ref{fig1}b shows a simplified \nv\ energy-level diagram as confirmed by experiment. The triplet-triplet (${^3}A_2\leftrightarrow{^3}E$) and singlet-singlet (${^1}E\leftrightarrow{^1}A_1$) energy differences are known to be 1.945 eV (637 nm) and 1.190 eV (1042 nm), respectively. \cite{davies_hamer, victor_singlets, manson_singlets, manson_singlets2} However, where these energy states lie with respect to the diamond valence and conduction bands is only known indirectly, as are the triplet-singlet (${^3}A_2\leftrightarrow{^1}A_1$ and ${^1}E\leftrightarrow{^3}E$) energy differences. \cite{wrachtrup_darkstates, toyli} Theoretical calculations predict the existence of additional energy states (${^1}E'$ and ${^1}A_1'$), but disagree on their energies (see Refs.~[\onlinecite{rand_lenef_theory, delaney_greer_larsson, manson_electronic_solns, maze_groups, sang_louie}] and references therein). Prior experiments and \textit{ab initio} calculations studied the phonon sidebands (PSBs) for the ${^3}A_2\rightarrow{^3}E$ and ${^3}E\rightarrow{^3}A_2$ transitions.\cite{davies_hamer, davies_vibr_spec, gali_NV_vibronics, zhang_vibronics, austr_thermo} The \sst\ and ${^1}A_1\rightarrow{^1}E$ PSBs have not been studied theoretically, and only the ${^1}A_1\rightarrow{^1}E$ transition had been measured prior to this work. \cite{victor_singlets, manson_singlets, manson_singlets2}

A more complete experimental picture of \nv\ properties can provide insight for applications and validate theoretical models of \nv\ attributes. The 1042 nm infrared \sst\ zero-phonon line (ZPL) has been used in an absorption-based magnetometer,\cite{victor_IRmag} but using the \sst\ PSB instead may be more sensitive depending on the PSB structure and cross section. In addition, most NV experiments take advantage of an optical pumping mechanism (which involves the $^1A_1$ excited vibrational states) that drives electrons to the $^3A_2$ $m_s=0$ state. Therefore, knowledge of the \sst\ PSB could improve infrared magnetometry and optical pumping schemes. Moreover, as the \nv\ center develops into a mature quantum system, it is important to know the properties of the singlet states to inspire confidence that we understand this system.

We attempt to fill the gaps in the knowledge of \nv\ properties by measuring the \sst\ PSB and searching for previously unobserved transitions. Finding the $^1E \rightarrow {^1}E'$ ZPL would resolve the disagreement on the predicted $^1E'$ energy. The \sst\ PSB yields information about the $^1A_1$ phonon modes, which are also of interest. The spin-orbit interaction mixes the $^3E$ and $^1A_1$ states, resulting in triplet-singlet intersystem crossing (ISC). This enables spin-dependent non-radiative decay from the nominally $^3E$ state to the nominally $^1A_1$ state. The ISC rate is comparable to the $^3E \rightarrow {^3}A_2$ spontaneous decay rate \cite{robledo_rates, tetienne_rates} and is an important factor in the optical pumping process. Measuring the $^1A_1$ phonon modes could allow the optical pumping mechanism to be modeled more accurately and provide insight on \nv\ spin polarization and readout. Furthermore, the accepted group-theoretical model of \nv\ predicts $^3A_2$ and $^1A_1$ to have the same electronic configuration, meaning they should have the same phonon modes. A comparison between the ${^3}E\rightarrow{^3}A_2$ and \sst\ PSBs should be sensitive to differences between the $^3A_2$ and $^1A_1$ configurations.

In this work, we present measurements of the \sst\ ZPL and PSB. We describe the PSB absorption features, including a high-energy (169 meV) localized phonon mode that lies outside the diamond lattice phonon density of states. Comparing the \sst\ and the ${^3}E\rightarrow{^3}A_2$ phonon modes, we find that the $^1A_1$ phonon modes are shifted to higher energies, meaning that proper descriptions of the $^1A_1$ and $^3A_2$ states require corrections to their electronic configurations.

\section{Experiment and Results}
\begin{figure}[h]
\includegraphics[width=0.48\textwidth]{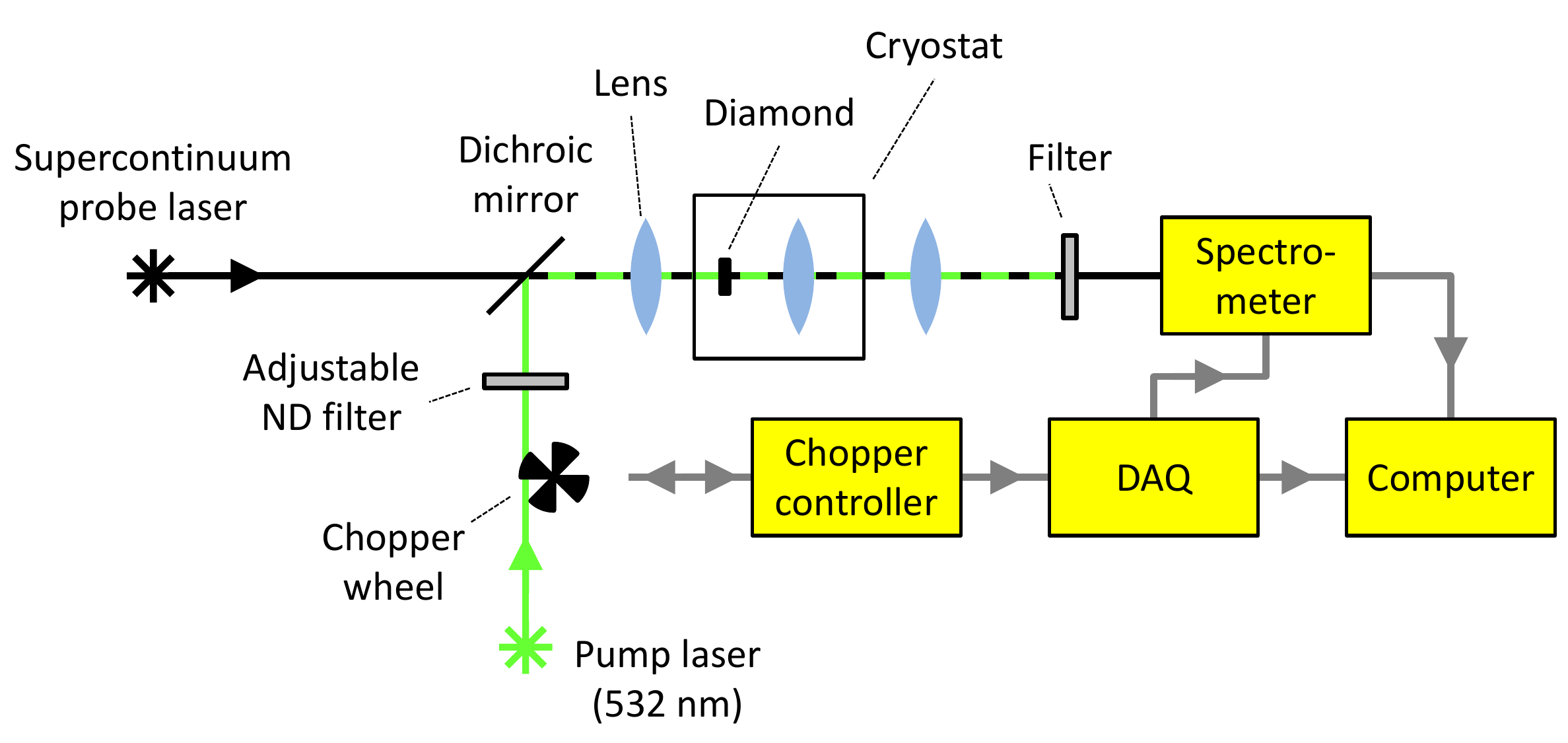}
\caption{\label{schem}The experimental apparatus. The data acquisition device (DAQ) monitors the chopper wheel state and triggers a spectrum acquisition when the pump is blocked or unblocked. The computer collects ``pump blocked" and ``pump unblocked" transmission spectra.}
\end{figure}

In our experiment, we populated the metastable ${^1}E$ state using pump-laser light and measured transmission of probe-laser light through a diamond sample containing an ensemble of \nv\ centers (Fig.~\ref{fig1}b and Fig.~\ref{schem}). We determined the probe transmission through the diamond with and without \nv\ centers in the ${^1}E$ state. A 532 nm frequency-doubled Nd:YVO$_4$ pump laser beam and a 5 mW supercontinuum probe laser beam (wavelength range 450-1800 nm) were combined on a dichroic beamsplitter and focused with a 40$\times$ microscope objective (0.6 numerical aperture) onto a cryogenically cooled diamond sample. The transmitted light was collimated and detected with a spectrometer with $\sim$1 nm resolution. A chopper wheel modulated the pump light and a computer collected a transmission spectrum each time the pump light was blocked and unblocked. Absorption from $^1E$ appeared as a difference between the ``pump blocked" and ``pump unblocked" supercontinuum transmitted intensities. In another experiment, we used 912 nm and 1042 nm continuous-wave (cw) lasers as probe sources and replaced the spectrometer with a photodiode.\cite{suppl}

\begin{figure}[h]
\begin{center}
\begin{overpic}[width=0.52\textwidth]{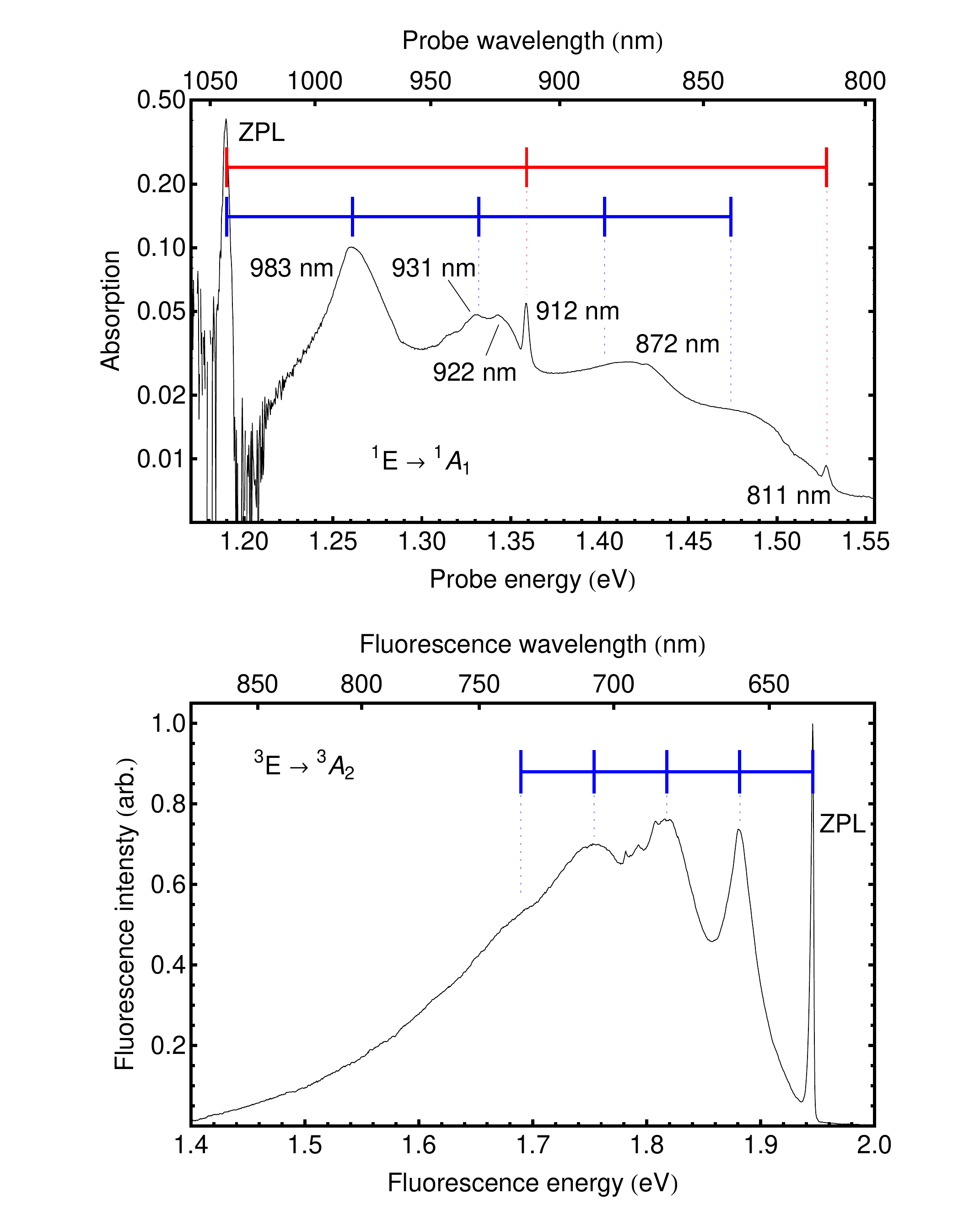}
\put(8,305){a}
\put(8,142){b}
\end{overpic}
\end{center}
\caption{\label{spec}(a) The supercontiuum absorption spectrum collected at 10 K for diamond sample B8 using 35 mW of pump-laser light. PSB fluorescence from $^3E \rightarrow {^3}A_2$ is present for wavelengths shorter than 840 nm and has been subtracted out. The vertical ticks indicate the expected PSB absorption energies for 71 and 169 meV phonons, which align with some of the absorption features. (b) The fluorescence spectrum of a similar diamond collected at 4 K. The vertical ticks indicate the expected PSB absorption energies for 64 meV phonons. Although the 686, 692, and 696 nm features are often ignored, they are vital to our comparison of the \sst\ and ${^3}E\rightarrow{^3}A_2$ PSBs, as they give rise to peaks (3)-(5) in Fig.~\ref{1phon}.}
\end{figure}

\begin{figure}[h]
\begin{center}
\includegraphics[width=0.48\textwidth]{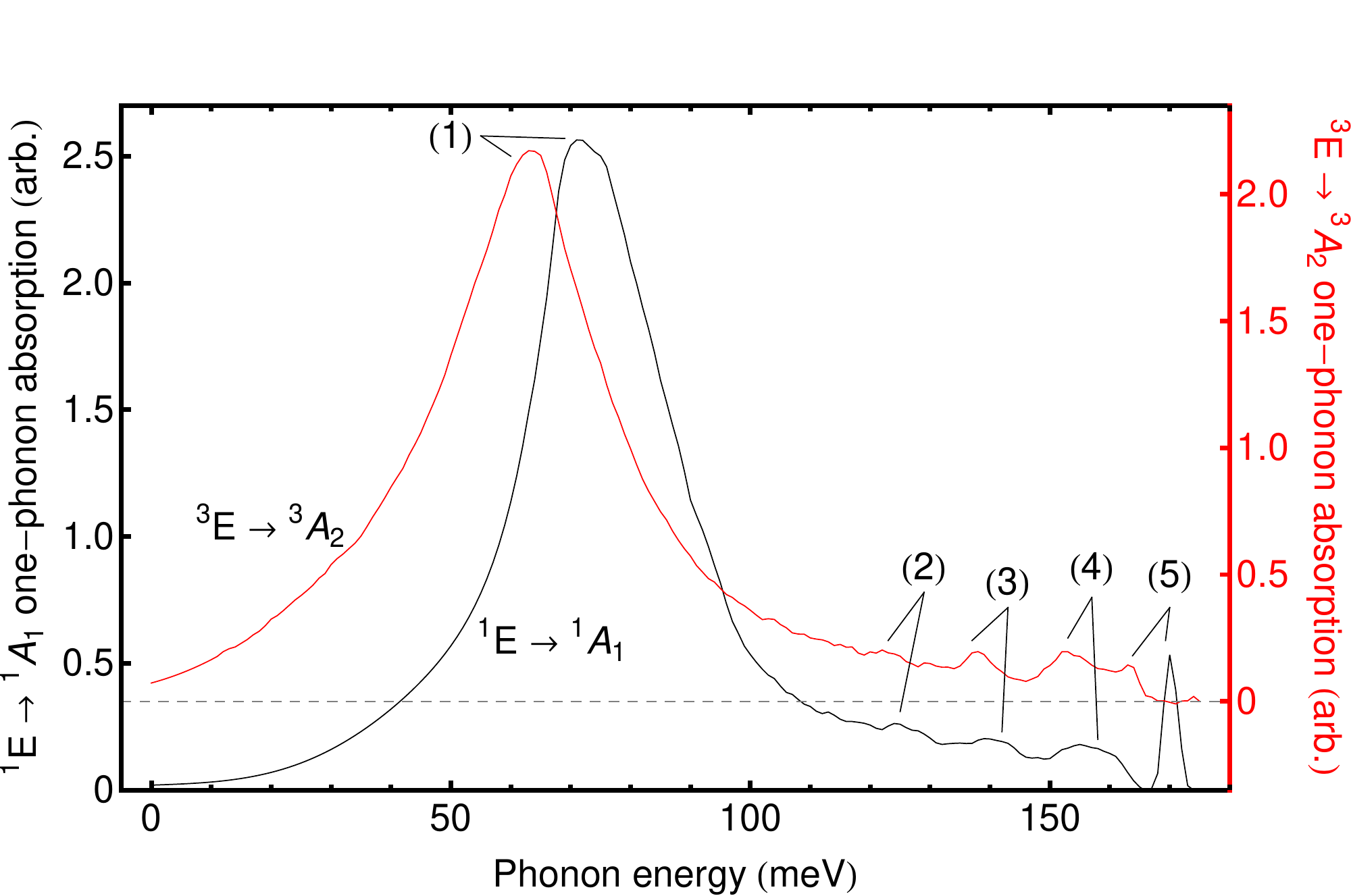}
\end{center}
\caption{\label{1phon}The one-phonon spectra for the \sst\ and ${^3}E\rightarrow{^3}A_2$ transitions, extracted from Fig.~\ref{spec}. The above spectra are normalized to have equal areas, and the $^3A_2$ curve is vertically offset for clarity. In each spectrum we see five peaks, labeled (1)-(5), though the $^1A_1$ peaks are shifted to higher energies (see Tab.~\ref{shiftTable}).}
\end{figure}

Figure \ref{spec}a shows the \sst\ ZPL and PSB supercontinuum absorption spectrum taken at 10 K with the sample ``B8", a synthetic type Ib high-pressure high-temperature (HPHT) diamond with $\sim$10 ppm \nv\ concentration. The PSB includes narrow absorption lines at 811 and 912 nm and broad absorption features at 872, 922, 931, and 983 nm. In the figure we observe that the 912 and 811 nm lines are 169 meV and 2$\times$169 meV away from the ZPL, respectively. Consequently, we believe the 811 and 912 nm lines are due to a 169.28(4) meV phonon mode and that the other lines are due to a distribution of phonon modes. Figure \ref{spec}b shows the ${^3}E\rightarrow{^3}A_2$ fluorescence spectrum taken at 4 K with a similar diamond (also $\sim$10 ppm \nv\ concentration). This PSB has a broader energy range, and has features at 686, 692, and 696 nm. Using these measured spectra and the techniques outlined in Refs.~[\onlinecite{davies_vibr_spec},  \onlinecite{suppl}, \onlinecite{maradudin}], we calculated the \sst\ and ${^3}E\rightarrow{^3}A_2$ Huang-Rhys parameters (0.9 and 3.49, respectively) as well as their one-phonon spectra (Fig.~\ref{1phon}), which are the rates at which these transitions create one phonon of a given energy. We expect these one-phonon spectra to be comparable, since both come from $E \rightarrow A$ transitions with similar final-state electronic configurations (Fig.~\ref{fig1}c). The one-phonon spectra show resemblance, and the differences between them are because of electronic Coulomb repulsion corrections to the $^1A_1$ level. These corrections mix the $^1A_1$ level with the higher-energy $^1A_1^\prime$ level. As a result, the $^1A_1$ level contains an admixture of configurations, which results in the difference in the one-phonon spectra.\cite{suppl}

We observed the above \sst\ PSB features in several diamond samples, and the absorption was greater in samples with higher \nv\ concentration. The \sst\ absorption should increase with pump power and saturate when the pumping rate becomes comparable to the $^1E$ decay rate. The absorption at room temperature increased linearly with pump power (up to 60 mW focused to a minimum beam waist smaller than 5 $\upmu$m), indicating that the $^1E$ population was not saturated. However, the absorption at 10 K saturated at $\sim$15 mW. This saturation is likely due to the prolonged $^1E$ lifetime at cryogenic temperature.\cite{victor_singlets} Introducing a static transverse magnetic field to the samples improved the absorption contrast by a few percent. This is because the Zeeman interaction mixes the triplet spin sublevels, which spoils the optical pumping to $m_s = 0$ and increases the $^1E$ population. 

We did not detect a $^1E \rightarrow{^1}E'$ ZPL in the 480-1100 nm range of the supercontinuum transmission spectrum, which means this transition lies outside of this range or was too weak to detect. This wavelength span was limited by the spectrometer.

Using a rate equation calculation based on the \nv\ excitation and decay rates at room temperature,\cite{tetienne_rates} we estimate the room temperature \sst\ ZPL cross section to be roughly $4\times 10^{-22}$ m$^2$, which is consistent with previous work. \cite{suppl, yannickIR} The accuracy of this cross section estimate is primarily limited by uncertainty in the \nv\ concentration; varying the \nv\ concentration from 5 to 20 ppm in our model yields estimated cross sections ranging from 3.4 to $5.4\times 10^{-22}$ m$^2$ (compared to $4.0\times 10^{-22}$ m$^2$ with 10 ppm \nv).

We varied the temperature of sample B8 from 10 to 300 K and recorded the absorption-feature contrasts, linewidths, and integrated areas.\cite{suppl} The features become weaker and broader with increasing temperature, and their integrated areas decrease. This decrease in area is consistent with the $^1E$ lifetime decrease observed in Ref.~[\onlinecite{victor_singlets}].

\begin{table}[h]
\begin{tabular}{r|r|r}
Peak \# & $^3A_2$ state & $^1A_1$ state \\
\cline{1-3}
(1) & 64 meV & 71 meV \\
(2) & 122 meV & 125 meV \\
(3) & 138 meV & 141 meV \\
(4) & 153 meV & 156 meV \\
(5) & 163 meV & 169 meV \\
\end{tabular}
\caption{\label{shiftTable} The energies of the one-phonon peaks shown in Fig.~\ref{1phon}. When comparing the energies of the $^3A_2$ and $^1A_1$ phonon modes, we see a systematic shift to higher energy of a few meV.}
\end{table}

Using cw probe lasers and a similar diamond sample ``S2" (16 ppm \nv\ concentration), we measured the center wavelengths of the 912 and 1042 nm absorption lines at 40 K to be 912.19(2) nm and 1041.96(2) nm. Our ZPL center wavelength is consistent with previous measurements.\cite{victor_singlets, manson_singlets} At low temperatures, the 1042 and 912 nm features have narrow widths (currently limited by the spectrometer resolution). These narrow widths imply that the vibrational mode associated with the 912 nm feature is sharp. By measuring 912 nm absorption as a function of light polarization angle, we found that the 912 nm absorption has the same polarization selection rules as the \sst\ ZPL.\cite{victor_singlets, suppl}

A 532 nm pump laser may excite other defects besides \nv\ (such as NV$^0$), meaning we must be cautious when associating the observed infrared absorption features with the \nv\ \sst\ transition. Selective excitation of infrared fluorescence using 637 nm pump light was shown in Ref.~[\onlinecite{manson_singlets2}], meaning that while the 1042 nm ZPL is surely related to \nv, we must convince ourselves that the other infrared absorption features are also part of this electronic transition. The one-phonon absorption spectrum (Fig.~\ref{1phon}), the optically-detected magnetic resonance test of the 912 nm selection rules \cite{suppl}, and the fact that a transverse magnetic field enhances the infrared absorption all confirm that our absorption spectrum belongs to the \nv\ \sst transition.

\section{Analysis and Discussion}
Comparing the \sst\ absorption PSB in Fig.~\ref{spec}a with previous observations of the $^1A_1\rightarrow{^1}E$  fluorescence PSB,\cite{manson_singlets, manson_singlets2} it is evident that these PSBs differ significantly. This difference is due to the anharmonicity of the $^1E$ vibronic levels induced by the dynamic Jahn-Teller effect, which is not present in $^1A_1$. \cite{manson_singlets2} In the low-temperature limit, the PSB features of $A\rightarrow E$ electronic transitions exhibit anharmonicity, while the PSB features of $E\rightarrow A$ transitions are harmonic.\cite{jtStudies2} Consequently, it is appropriate to compare the ${^1}E\rightarrow{^1}A_1$ absorption PSB with the ${^3}E\rightarrow{^3}A_2$ fluorescence PSB. Furthermore, $^1A_1$ and $^3A_2$ have the same electronic configuration ($a_1^2e^2$) when electronic Coulomb repulsion is ignored, meaning they should have similar nuclear equilibrium positions and phonon modes. Since their initial states are different, the \sst\ and ${^3}E\rightarrow{^3}A_2$ transitions may couple to a different number of phonon modes and have different Huang-Rhys parameters, but the $^1A_1$ and $^3A_2$ one-phonon spectra should be similar.

As mentioned above, we extracted the one-phonon spectra from the PSBs shown in Fig.~\ref{spec}. The \textit{n}-phonon spectrum is the convolution of the (\textit{n}-1)-phonon and one-phonon spectra, and the sum of all \textit{n}-phonon spectra generates the transition PSB. The one-phonon spectra are also related to the $^1A_1$ and $^3A_2$ phonon density of states (DOS). As seen in Fig.~\ref{1phon}, we found similarities between the one-phonon spectra; both spectra have one large feature and four small features. However, all of the \sst\ features are displaced to higher energies (Tab.~\ref{shiftTable}). 

Introducing a point defect into a lattice alters the vibrational motion of the defect and its neighbors from what it would have been with ordinary atoms in the lattice. This is because the parameters that determine the frequencies of the vibrational motion for these atoms (the masses and effective spring constants) are modified. When the frequencies of the local oscillations of the defect lie within the spectrum of allowed vibrational modes of the remaining crystal, the local modes hybridize with the lattice modes and are called ``quasilocal" (quasilocal because the nuclear oscillation amplitudes fall off slowly with increasing distance from the defect). \cite{maradudin, zhang_vibronics} The $\sim$71 meV phonon modes we observed appear to be from a quasilocal mode of \nv\ in the $^1A_1$ state. The diamond lattice phonon DOS is appreciable at 71 meV,\cite{lattice_dos3, lattice_dos_expt} and since the \nv\ 71 meV mode couples strongly to the diamond lattice modes, the peaks of the 71 meV mode are consequently broadened.

In contrast to the quasilocal mode case, a ``localized" mode occurs when the frequency of the local oscillations of a defect lies outside the lattice phonon DOS. In this instance, the oscillations of the defect couple poorly to the oscillations of the rest of the crystal, the vibrational motion is confined to the region of the defect, and the local phonon mode energy is unbroadened.  This is the case for the 169 meV mode. The diamond lattice phonon DOS has an upper limit of 168 meV.\cite{lattice_dos3, lattice_dos_expt, zaitsev} The \nv\ 169 meV mode falls outside the diamond lattice phonon spectrum and couples poorly to the lattice modes, consequently making the peaks of the 169 meV mode in Fig.~\ref{spec}a sharp.

The existence of a 169 meV local phonon mode and the differences between the \sst\ and ${^3}E\rightarrow{^3}A_2$ one-phonon spectra are surprising for several reasons. \textit{Ab initio} calculations for the \nv\ triplet-state vibrations do not predict the existence of high-energy local phonon modes,\cite{gali_NV_vibronics, zhang_vibronics} and the \sst\ PSB is the only \nv\ PSB to contain such a feature. Due to the discrepancy in one-phonon spectra, we conclude that the $^1A_1$ level has electronic Coulomb repulsion corrections that modify its phonon modes from those of the $^3A_2$ level. Since the features in the one-phonon spectrum are shifted to higher energies, we can determine that the nearby atoms are more tightly bonded in the $^1A_1$ level than in the $^3A_2$ level.

\section{Outlook}
In summary, we measured the \sst\ absorption spectrum of the \nv\ center using pump-probe spectroscopy. In the \sst\ PSB and one-phonon absorption spectrum we found several phonon modes, one of which lies outside the diamond lattice phonon DOS. The \sst\ and ${^3}E\rightarrow{^3}A_2$ one-phonon spectra show general similarity, but the $^1A_1$ phonon modes are shifted to higher energies, which is from corrections to the $^1A_1$ orbital configuration due to electronic Coulomb repulsion (not included in other theories). Our measurement of the \sst\ absorption spectrum shows that the ZPL is more absorptive than the PSB, and hence the ZPL offers greater sensitivity for infrared-absorption-based magnetometry than the PSB wavelengths. Furthermore, the \nv\ ISC and optical pumping process can be modeled more precisely using our measured $^1A_1$ vibronic structure.

We searched for the ${^1}E\rightarrow{^1}E'$ ZPL for energies up to 2.0 eV at cryogenic temperature and 2.6 eV at room temperature, but we did not detect it. The ${^3}A_2\rightarrow{^3}E$ and ${^1}E\rightarrow{^1}E'$ should have similar cross sections because they are transitions from electronic configuration $a_1^2 e^2$ to $a_1 e^3$ (neglecting Coulomb coupling). Since the \sst\ ZPL cross section is smaller than that of ${^3}A_2\rightarrow{^3}E$ (see Ref.~[\onlinecite{wee_2photon}]), the ${^1}E\rightarrow{^1}E'$ transition should have a similar or larger cross section compared to the \sst\ transition. This means the ${^1}E\rightarrow{^1}E'$ ZPL would likely have been detected in our absorption measurements if its energy is less than 2.0 eV. This suggests that the ${^1}E\rightarrow{^1}E'$ ZPL energy is greater than 2.0 eV. Follow-up experiments will extend the search for the ${^1}E\rightarrow{^1}E'$ ZPL to higher energies with improved sensitivity.

\begin{acknowledgements}
We are grateful to the group of Prof.~F.~Wang (UC Berkeley) for help with the supercontinuum laser. We thank V.~Acosta and C.~Santori (Hewlett-Packard Laboratories), and V.~Huxter and S.~Choi (UC Berkeley) for useful discussions. This work was supported by the NSF, DOE SCGF, the AFOSR/DARPA QuASAR program, NATO SFP, IMOD, ARC (DP120102232), and the Danish Council for Independent Research in Natural Sciences.
\end{acknowledgements}
\bibliography{bibl_forsuppl}

\onecolumngrid
\newpage


\section*{Supplementary material}

# Supplementary material to "The Infrared Absorption Band and Vibronic Structure of the Nitrogen-Vacancy Center in Diamond"

P. Kehayias, M.W. Doherty, D. English, R. Fischer, A. Jarmola,

K. Jensen, N. Leefer, P. Hemmer, N.B. Manson, and D. Budker

## INTRODUCTION

In this supplement we present additional content to the results described in the main article. We include additional experimental technical details and findings, a description of how we calculated the one-phonon spectra from the $^1E \rightarrow {}^1A_1$ and $^3E \rightarrow {}^3A_2$ absorption and emission PSBs, and an analysis of the resulting $^3A_2$ and $^1A_1$ vibronic bands.

## EXPERIMENTAL DETAILS

We used an Olympus LUCPlanFL N 40× PH2 microscope objective (0.6 numerical aperture) to focus the pump and probe beams onto the diamond samples (to a minimum beam waist smaller than 5 µm). This objective lens is achromatic, which ensures that the pumped NV$^-$ and probe spatial regions overlap. We used a Janis ST-500 liquid-helium flow cryostat for cooling the diamond sample. Our pump-laser source was a Coherent Verdi-V6, and our cw probe-laser sources included 2 mW of 912 nm light from a Coherent CR 899 Ti:Sapphire laser, 30 mW of 912 nm light from a diode laser (1 nm linewidth), and 1.5 mW of 1042 nm light from an external-cavity diode laser (ECDL). We used a Fianium SC450-2 supercontinuum laser as our broadband probe and an Ocean Optics USB2000+VIS-NIR (~1 nm resolution, optimized for infrared sensing) for detecting transmitted supercontinuum light.

We used the $^3E \rightarrow {}^3A_2$ fluorescence spectrum of a similar diamond sample at 4 K for comparison with our $^1E \rightarrow {}^1A_1$ spectrum. This sample was also illuminated with 532 nm pump light, and the emitted fluorescence was dispersed with a monochromator (0.1 nm resolution) and detected with a photomultiplier with a GaAs photocathode (calibrated with the blackbody spectrum from a 3100 K tungsten bulb). Although we compared spectra from different diamond samples, the spectra from different high-concentration samples are

in general consistent.

## $^1E \rightarrow {}^1A_1$ ZPL CROSS SECTION ESTIMATE

Using experimental decay-rate and pump-laser absorption cross section parameters [2, 3], we constructed a rate-equation model to estimate the fraction of NV$^-$ centers in the metastable state throughout the diamond sample, from which we estimated the $^1E \rightarrow {}^1A_1$ ZPL cross section at room temperature. Since the Rayleigh range is about 150 µm (compared to a 730 µm thick diamond sample) and the 532 nm optical depth is about 4, our model takes into account the pump beam divergence and absorption in the diamond. Using this calculation of the metastable NV$^-$ center density and the experimentally determined 1042 nm transmission in sample B8 at various pump powers, we determined the $^1E \rightarrow {}^1A_1$ ZPL cross section. For simplicity, we approximated the probe beam to be a straight line through the pump-beam axis. We estimated the $^1E \rightarrow {}^1A_1$ ZPL cross section to be $4 \times 10^{-22}$ m$^2$. This value is consistent with that of Ref. [4]. As mentioned in the main text, the uncertainty in NV$^-$ center concentration dominates the error on this cross section estimate. Uncertainty in the NV$^-$ center excited-state decay rates [2], the pump-laser absorption cross section [3], the pump beam waist, and the distance between the pump beam focus and the diamond surface contribute an additional uncertainty of about $0.9 \times 10^{-22}$ m$^2$.

## TEMPERATURE DEPENDENCE OF THE $^1E \rightarrow {}^1A_1$ ZPL AND PSB

We measured $^1E \rightarrow {}^1A_1$ supercontinuum absorption spectra while varying the temperature of sample B8 from 10 to 300 K. Figure S1 shows the absorption, linewidths, and integrated areas for the 1042 nm ZPL and the 811, 912, and 983 nm PSB features. These features became weaker and broader with increasing temperature, and the integrated areas decreased. The integrated areas should be independent of temperature [5]. However, the $^1E$ state has shorter lifetime at higher temperature due to the enhanced electron-phonon decay rate to $^3A_2$ [6]. We believe the decrease in integrated area is because of the consequent reduction in $^1E$ population at higher temperature.

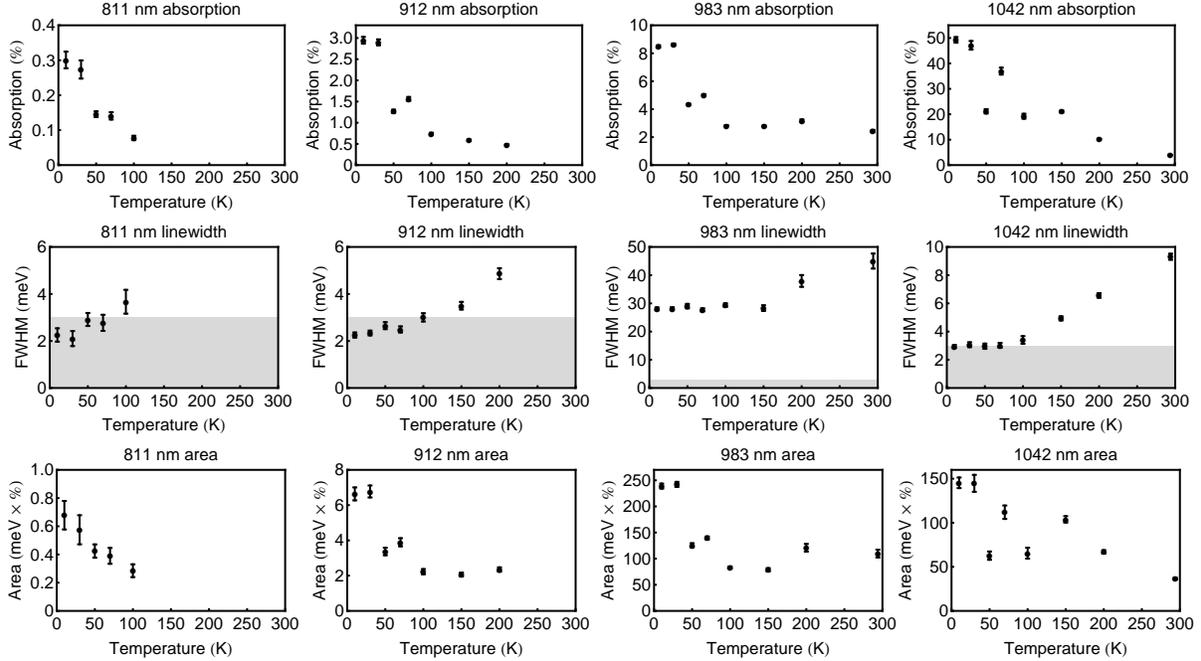

FIG. S1. The percent absorption, linewidth, and integrated area of $^1E \rightarrow {}^1A_1$ absorption features in sample B8 as a function of temperature. The 811 and 912 nm lines vanished in the supercontinuum absorption spectra above 100 and 200 K, respectively. The spectrometer resolution contributes to the apparent linewidths plotted above; the gray region (3 meV) indicates where the spectrometer instrument broadening significantly contributes to the measured linewidths. We believe the 50 K and 90 K measurements to be outliers due to thermal expansion in the cryostat during the measurement. The above error bars are one-sigma statistical errors extracted from the parameter fits of the absorption spectra. We estimate a 1 meV systematic uncertainty on the above linewidths.

## 912 NM POLARIZATION SELECTION RULES

We investigated the light-polarization selection rules for 912 nm absorption and compared them to those listed in Table S1. An $E \rightarrow A_1$ transition is dipole-allowed for $(x, y)$-polarized light, while an $E \rightarrow E$ transition is also dipole-allowed for $z$-polarized light. A difference between the 912 nm selection rules and the expected $E \rightarrow A_1$ ZPL selection rules could indicate that the 912 nm line is an $E \rightarrow E$ transition or that the $^1E \rightarrow {}^1A_1$ selection rules are not strictly obeyed in PSB transitions [6]. We determined the polarization dependence of 912 nm absorption in a room-temperature optically-detected magnetic resonance (ODMR) experiment with diamond sample S2. We singled out the [111]-oriented NV$^-$ centers with an axial 15 G static magnetic field, exposed the sample to microwaves from a nearby wire, and measured the diode-laser absorption as a function of microwave frequency. Microwaves

resonant with $^3A_2$ $m_s = 0 \rightarrow m_s = \pm 1$ transitions spoil the optical spin polarization, increase $^1E$ population, and enhance probe absorption. By measuring the $m_s = 0 \rightarrow m_s = \pm 1$ absorption in [111]-oriented centers as a function of polarization angle for probe-light wavevector $\hat{k}$ parallel and perpendicular to the [111] $z$-axis, we found that the 912 nm transition is $(x, y)$-allowed and $z$-forbidden (Fig. S2). These selection rules indicate an $E \rightarrow A_1$ transition and are consistent with the $^1E \rightarrow ^1A_1$ ZPL selection rules [6]. We performed this experiment at 40 K with the 912 nm Ti:Sapphire laser and obtained consistent results for $\hat{k} \parallel z$, but we were unable to test $\hat{k} \perp z$ because of mechanical constraints.

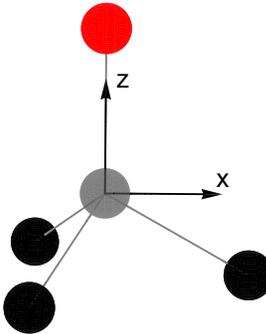

| Transition | Dipole-allowed |
|---|---|
| $A_1 \leftrightarrow A_1$ | $z$ |
| $A_1 \leftrightarrow A_2$ | — |
| $A_1 \leftrightarrow E$ | $(x, y)$ |
| $A_2 \leftrightarrow A_2$ | $z$ |
| $A_2 \leftrightarrow E$ | $(x, y)$ |
| $E \leftrightarrow E$ | $(x, y), z$ |

TABLE S1. Photon polarizations for dipole-allowed transitions between $C_{3v}$ electronic states [5]. The notation "$(x, y)$" implies that any polarization in the $x$-$y$ plane has the same transition amplitude. The drawing on the right indicates the choice of axes.

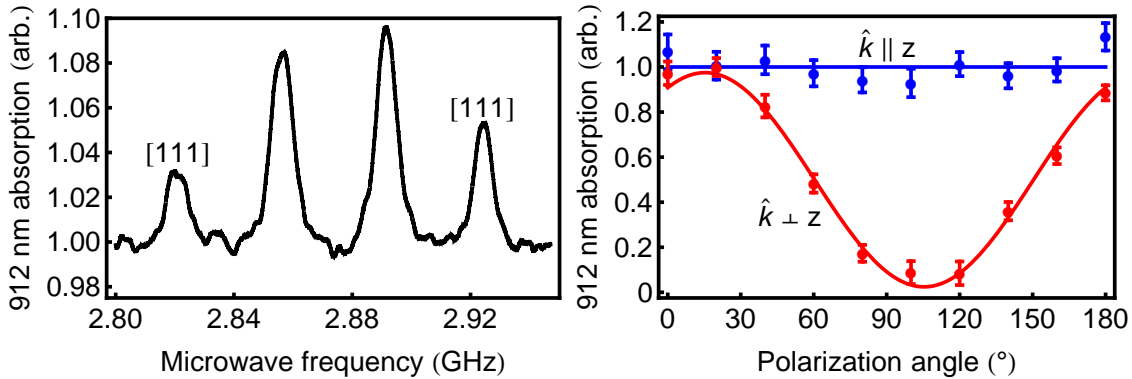

FIG. S2. (Left) A sample ODMR spectrum ($\hat{k} \perp z$) taken at room temperature, with absorption peaks for [111]-oriented centers labeled. The middle two peaks are due to the other three orientations. (Right) 912 nm absorption by [111]-oriented centers at different light polarization angles. The $\hat{k} \parallel z$ case is only sensitive to $x, y$ polarization while the $\hat{k} \perp z$ case is also sensitive to $z$ polarization. The constant nonzero $\hat{k} \parallel z$ ODMR peak height and sinusoidal variation to zero in the $\hat{k} \perp z$ data indicate the transition is $(x, y)$-allowed and $z$-forbidden.

## VIBRONIC ANALYSIS

### The vibronic structure predicted by the current NV$^-$ model

By adopting the adiabatic and harmonic approximations, the nuclear vibrational motion associated with a given electronic state is approximately governed by the harmonic potential formed by the dependence of the state's electronic energy on the nuclear coordinates. The nuclear vibrational potential associated with the $n^{th}$ electronic state $|\Phi_n\rangle$ is [7]

$$E_n(\vec{Q}) = E_n(0) + \sum_{\alpha,j,k} a_{n,\alpha,j,k} Q_{\alpha,j,k} + \frac{1}{2}(\omega_{\alpha,j,k}^2 + b_{n,\alpha,j,k})Q_{\alpha,j,k}^2 \qquad (1)$$

$$= E_n(0) - \delta E_n(0) + \frac{1}{2}\sum_{\alpha,j,k}(\omega_{\alpha,j,k}^2 + b_{n,\alpha,j,k})(Q_{\alpha,j,k} - \delta Q_{n,\alpha,j,k})^2, \qquad (2)$$

where

$$E_n(0) = \langle\Phi_n|\,\hat{H}_e\,|\Phi_n\rangle$$

$$a_{n,\alpha,j,k} = \langle\Phi_n|\,\frac{\partial\hat{H}_e}{\partial Q_{\alpha,j,k}}\bigg|_0\,|\Phi_n\rangle$$

$$\omega_{\alpha,j,k}^2 = \langle\Phi_0|\,\frac{\partial^2\hat{H}_e}{\partial Q_{\alpha,j,k}^2}\bigg|_0\,|\Phi_0\rangle$$

$$b_{n,\alpha,j,k} = \langle\Phi_n|\,\frac{\partial^2\hat{H}_e}{\partial Q_{\alpha,j,k}^2}\bigg|_0\,|\Phi_n\rangle - \omega_{\alpha,j,k}^2$$

$$\delta E_n(0) = \sum_{\alpha,j,k}\frac{a_{n,\alpha,j,k}^2}{2(\omega_{\alpha,j,k}^2 + b_{n,\alpha,j,k})}$$

$$\delta Q_{n,\alpha,j,k} = -\frac{a_{n,\alpha,j,k}}{2(\omega_{\alpha,j,k}^2 + b_{n,\alpha,j,k})}. \qquad (3)$$

In the above expressions, $\hat{H}_e$ is the electronic Hamiltonian, $Q_{\alpha,j,k}$ is the normal nuclear displacement coordinate (with respect to the nuclear equilibrium coordinates of the ground $n = 0$ electronic state) of the $\alpha^{th}$ ground electronic state eigenmode with symmetry $(j,k) = \{A_1,\ A_2,\ (E,x),\ (E,y)\}$ and vibrational energy $\hbar\omega_{\alpha,j,k}$. Note that quadratic terms that couple the eigenmodes in excited electronic states have been ignored and that for the ground electronic state $a_{0,\alpha,j,k} = 0$ is a condition of nuclear equilibrium. It is clear that linear interactions displace the nuclear equilibrium coordinates by $\delta Q_{n,\alpha,j,k}$ and that the quadratic interactions shift the vibrational energies by $\delta\omega_{n,\alpha,j,k} \approx b_{n,\alpha,j,k}/2$.

The vibrational state $|\chi_{n,\nu}\rangle$ associated with the $n^{th}$ electronic state with vibronic energy

$E_{n,\nu}$ follows trivially from the vibrational equation $[\hat{T}(\vec{Q}) + E_n(\vec{Q})]\,|\chi_{n,\nu}\rangle = E_{n,\nu}\,|\chi_{n,\nu}\rangle$, where $\hat{T}(\vec{Q})$ is the nuclear kinetic energy. Importantly, the vibronic coupling of the adiabatic vibronic states $|\Phi_n, \chi_{n,\nu}\rangle$ by Jahn-Teller interactions between degenerate electronic states has not been considered thus far. The Jahn-Teller interactions occur between degenerate electronic states and degenerate phonon modes and result in anharmonicities in the vibronic structures of the degenerate electronic states. For the case of the NV$^-$ center, only $E$-symmetric electronic states and phonon modes exhibit degeneracy. As discussed in the article, the vibronic couplings and anharmonicites induced by the Jahn-Teller effect are not relevant to the PSBs analyzed in this work.

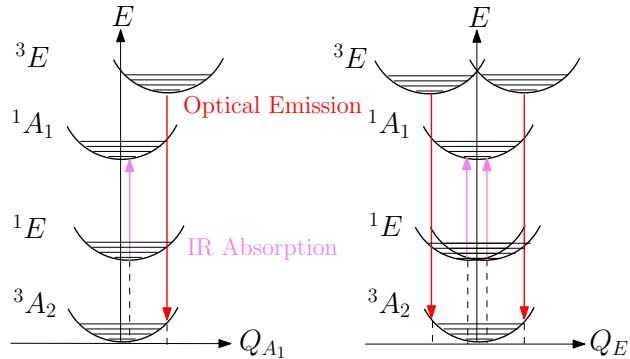

FIG. S3. Configuration coordinate diagrams for $A_1$ ($Q_{A_1}$) and $E$ ($Q_E$) phonon modes depicting the harmonic nuclear potential wells and phonon energy levels. The optical emission and IR absorption transitions (solid arrows) are depicted according to the Franck-Condon principle. Previous electronic structure models predict that the wells of the $^3E$ and $^1E$ are displaced from the $^3A_2$ equilibrium coordinates (as indicated by dashed lines) differently, but the well of $^1A_1$ is not displaced. The displacements of the degenerate electronic levels of $^3E$ and $^1E$ are equal and opposite (displacements do not imply static Jahn-Teller effects). Previous electronic models also predict that the phonon energies of $^3E$ and $^1E$, but not $^1A_1$, may differ from those of $^3A_2$.

To elaborate on this discussion, the IR absorption and optical emission PSBs are both $E \to A$ electronic transitions and thus have analogous configuration coordinate diagrams (refer to Fig. S3). At low temperatures where only the ground vibronic levels of $^3E$ and $^1E$ are populated, the features of the PSBs occur at the harmonic vibronic energies of $^3A_2$ and $^1A_1$ and are unaffected by the Jahn-Teller anharmonicities of $^3E$ and $^1E$. Furthermore, for typical Jahn-Teller interactions, the contributions of $E$ modes to PSBs of $E \to A$ electronic transitions are well described by the techniques applied to model $A_1$ mode contributions [8]. It follows that a comparison of the PSBs can relate the vibronic structures of $^3A_2$ and $^1A_1$ and also the electron-phonon interactions that give rise to the PSBs.

In the low-temperature limit, the Huang-Rhys parameter of a transition between the ground and $n^{th}$ excited electronic states is given by [7, 9]

$$S_n = \sum_{\alpha,j,k} \frac{\delta E_{n,\alpha,j,k}}{\hbar\omega_{n,\alpha,j,k}} \approx \int_0^\Omega S_n(\omega)\rho(\omega)\,d\omega, \tag{4}$$

where $\rho(\omega)$ is the density of modes of the ground electronic state, $\Omega$ is the highest mode frequency, and (ignoring mode energy shifts) $S_n(\omega) = \overline{a_n^2}(\omega)/2\hbar\omega^3$, such that $\overline{a_n^2}(\omega)$ is the average squared linear interaction parameter of all modes with frequency $\omega$. Likewise, the bandshape function due to linear interactions only is [7, 9]

$$g_n(\omega) = N_n S_n(\omega)\rho(\omega), \tag{5}$$

where $N_n$ is a normalization constant satisfying $N_n \int_0^\Omega g_n(\omega)\,d\omega = 1$. Note the minor differences in the above expressions to those that appear in Refs. [7] and [9].

Using the well-established expressions of the NV$^-$ electronic states in terms of molecular orbitals (MOs) ($a_1$, $e_x$ and $e_y$) [10], the linear and quadratic interaction parameters may be derived for each of the NV$^-$ center's electronic levels correct to second-order in electron-electron Coulomb repulsion (refer to Table S2). Note that in earlier models of the NV$^-$ center only the $^1E$ and $^1E'$ levels have been considered to couple by Coulomb repulsion [10, 11]. Here we also consider the Coulomb coupling of the $^1A_1$ and $^1A_1'$ levels. The Coulomb coupling of the $^1E$ and $^1A_1$ levels can be described by the parameters $\kappa_E$ (denoted $\kappa$ in previous work [10]) and $\kappa_A$, respectively. Table S2 demonstrates that unless the Coulomb coupling of the $A_1$ singlets is included, the current theoretical model predicts that the vibrational parameters of $^1A_1$ do not differ from those of $^3A_2$, whereas the parameters do differ at zero-order for $^3E$ and at first- (linear interactions with $E$ modes) and second-order (linear interactions with $A_1$ modes and all quadratic interactions) in $\kappa_E$ for $^1E$. If the Coulomb coupling of the $A_1$ singlet levels is included, then the vibrational parameters of $^1A_1$ differ from those of $^3A_2$ at second-order in $\kappa_A$ for linear interactions with $A_1$ modes and quadratic interactions with both $A_1$ and $E$ modes. Thus, including the Coulomb coupling of the $A_1$ singlets, the $A_1$ and $E$ modes contribute linearly at zero-order to the $S$ and $g(\omega)$ of the NV$^-$ optical emission PSB, but that $A_1$ and $E$ modes contribute linearly at different orders in $\kappa_A$ and $\kappa_E$ to the IR absorption PSB. Furthermore, including the Coulomb coupling of the $A_1$ singlets means

that the mode energies of the $^1A_1$ and $^1E$ levels differ from the $^3A_2$ level at second-order in $\kappa_A$ and $\kappa_E$, respectively.

In summary, the phonon mode energies of the $^1A_1$ and $^3A_2$ levels should differ slightly due to the second-order Coulomb coupling of the $A_1$ singlets. The contributions of $A_1$ and $E$ modes to the $^3E \to {}^3A_2$ and $^1E \to {}^1A_1$ bandshape functions should vary, resulting in different PSB features. Our measurements yield the same features in both the $^3E \to {}^3A_2$ and $^1E \to {}^1A_1$ bandshape functions, though they are shifted to higher energies in the latter. There is little evidence that the contributions of $A_1$ and $E$ modes to the $^1E \to {}^1A_1$ bandshape function result in different features to those of the $^3E \to {}^3A_2$ bandshape function.

### Decomposition and analysis of the vibronic bands

Adopting the well-established techniques introduced by Maradudin [12] and applied extensively by Davies [9] to color centers in diamond, the normalized PSB $I(\omega)$ of an electronic transition can be described by

$$I(\omega) = e^{-S} \int_{-\infty}^{\infty} e^{-\gamma|t|} e^{Sg(t)} e^{i\omega t}\, dt$$
$$= e^{-S} I_0(\omega) + e^{-S} \sum_{n=1}^{\infty} \frac{S^n}{n!} I_0 \otimes I_n(\omega), \tag{6}$$

where $S$ is the Huang-Rhys parameter, $\gamma$ is the homogeneous ZPL width, $I_0$ is the homogeneous ZPL shape (as parameterized by $\gamma$), $g(t) = (1/2\pi) \int_{-\Omega}^{\Omega} g(\omega) e^{-i\omega t}\, d\omega$ is the bandshape function, $I_1(\omega) = g(\omega)$ if $-\Omega \leq \omega \leq \Omega$ and $= 0$ otherwise, $I_n = I_1 \otimes I_{n-1}(\omega)$ for $n > 1$, $\Omega$ is the highest phonon frequency of diamond, and $\otimes$ denotes the operation of convolution. Note that the individual band components $I_n(\omega)$ offer the insightful interpretation of being representative of all processes of net energy $\hbar\omega$ involving the creation and/or annihilation of $n$ phonons. The total PSB is thus the summation over all $n$-phonon processes.

Absorption $I_{\text{abs.}}(\omega)$ and emission $I_{\text{em.}}(\omega)$ spectra are related to the normalized PSB via

$$\frac{I_{\text{abs.}}(\omega_0 + \omega)}{\omega_0 + \omega} \propto \frac{I_{\text{em.}}(\omega_0 - \omega)}{(\omega_0 - \omega)^3} \propto I(\omega), \tag{7}$$

where $\hbar\omega_0$ is the ZPL energy. The normalization condition $\int_{-\infty}^{\infty} I(\omega)\, d\omega = 1$ ensures that knowledge of the proportionality factors is not required to obtain the normalized PSB from

absorption/emission spectra. Once normalized, the Huang-Rhys parameter can be evaluated using $S = -\log\langle I_0\rangle$, where $\langle I_0\rangle$ is the integrated area of the ZPL. By applying Fourier techniques, the following expression for the bandshape function can be derived

$$g(\omega) = \frac{1}{S}\int_{-\infty}^{\infty}\log\{e^S e^{\gamma|t|}\int_{-\infty}^{\infty}[I(\omega) - e^{-S}I_0(\omega)]e^{-i\omega t}\,d\omega + 1\}e^{i\omega t}\,dt. \tag{8}$$

The above expression was found to be non-trivial to evaluate numerically and particularly sensitive to spectral noise. Consequently, it could not be used directly to obtain the optical emission and IR absorption bandshapes. After smoothing the spectra and evaluating approximate bandshape functions using the above expression, the evaluated bandshape functions were then used as seed functions to the iterative process first described by Mostoller *et al.* [13]. The process involves calculating each of the normalized PSB components $I_n(\omega)$ via convolutions of a seed $I_1(\omega)$ and then using

$$I_1(\omega) = e^S I(\omega) - I_0(\omega) - \sum_{n=2}^{\infty}\frac{S^n}{n!}I_0 \otimes I_n(\omega) \tag{9}$$

to obtain an improved estimate of $I_1(\omega)$. If a reasonable seed bandshape function was obtained directly from (8), we found that the iterative process converged within a few cycles.

The fitted IR absorption and optical emission PSBs, together with their individual band components $I_n(\omega)$, are depicted in Fig. S4 and Fig. S5, respectively. The plots demonstrate that the calculated PSBs reproduce all of the key features of the observed PSBs, but minor differences appear at higher energy. These differences are more pronounced for the IR absorption PSB and are likely to be related to anharmonicity introduced by higher-order vibronic effects. Further investigation is required. As concluded in the main text, the features of the optical emission and IR absorption PSBs result from electron-phonon interactions with similar phonon modes and that the mode energies are greater in $^1A_1$ than in $^3A_2$. The upward energy shift of the phonon modes in the $^1A_1$ level results in a local mode appearing with an energy greater than those of the diamond lattice.

**Analysis of the PSBs via a comparison with the phonons of diamond**

Having established the relationships between the PSBs, it is left to explain the origins of the PSB features. If we assume that in $^3A_2$, the NV$^-$ center only slightly perturbs the

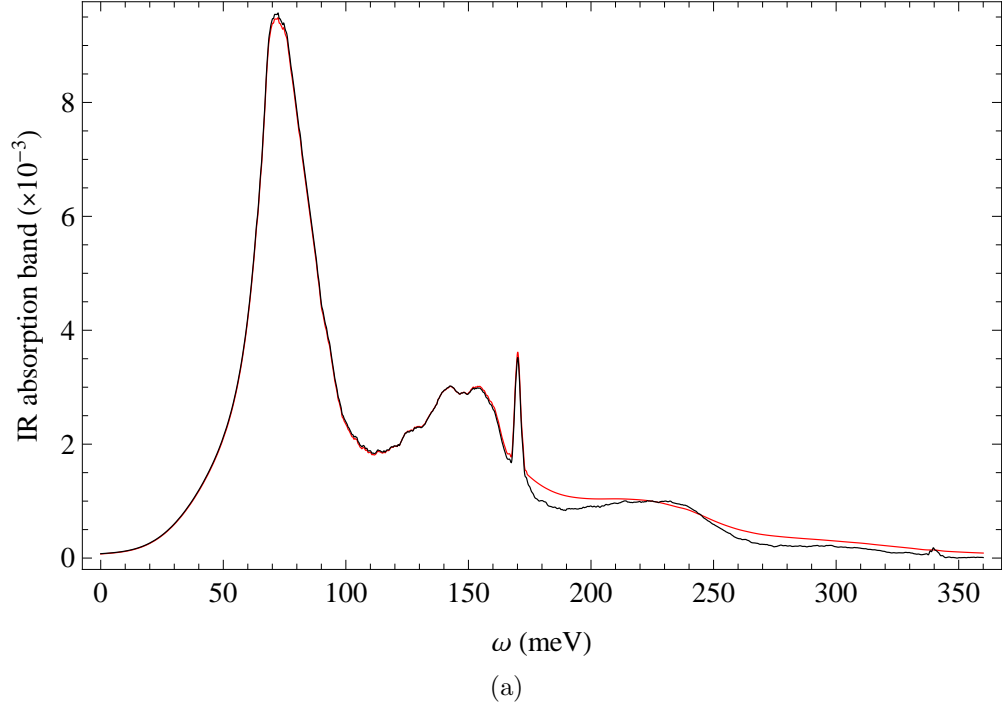

(a)

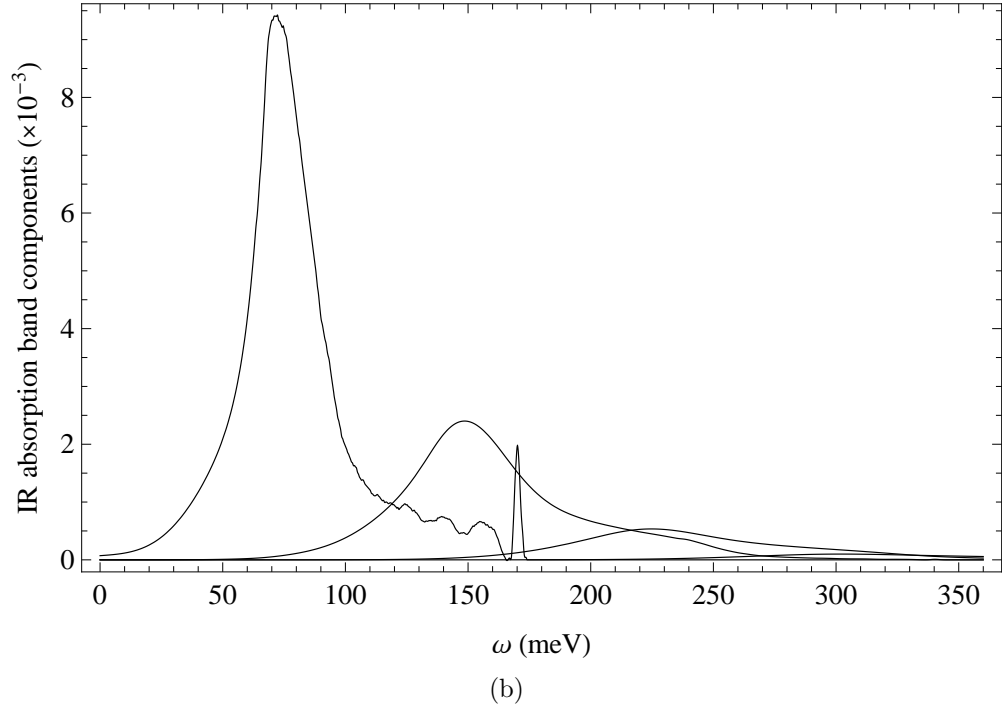

(b)

FIG. S4. (a) The normalized observed IR absorption PSB (black) and the calculated PSB (red) obtained using the IR absorption $g(\omega)$ and (6). (b) The calculated $n$-phonon components (increasing from $n$=1 from left to right) of the IR absorption PSB. The sum of the components equals the calculated PSB depicted in (a).

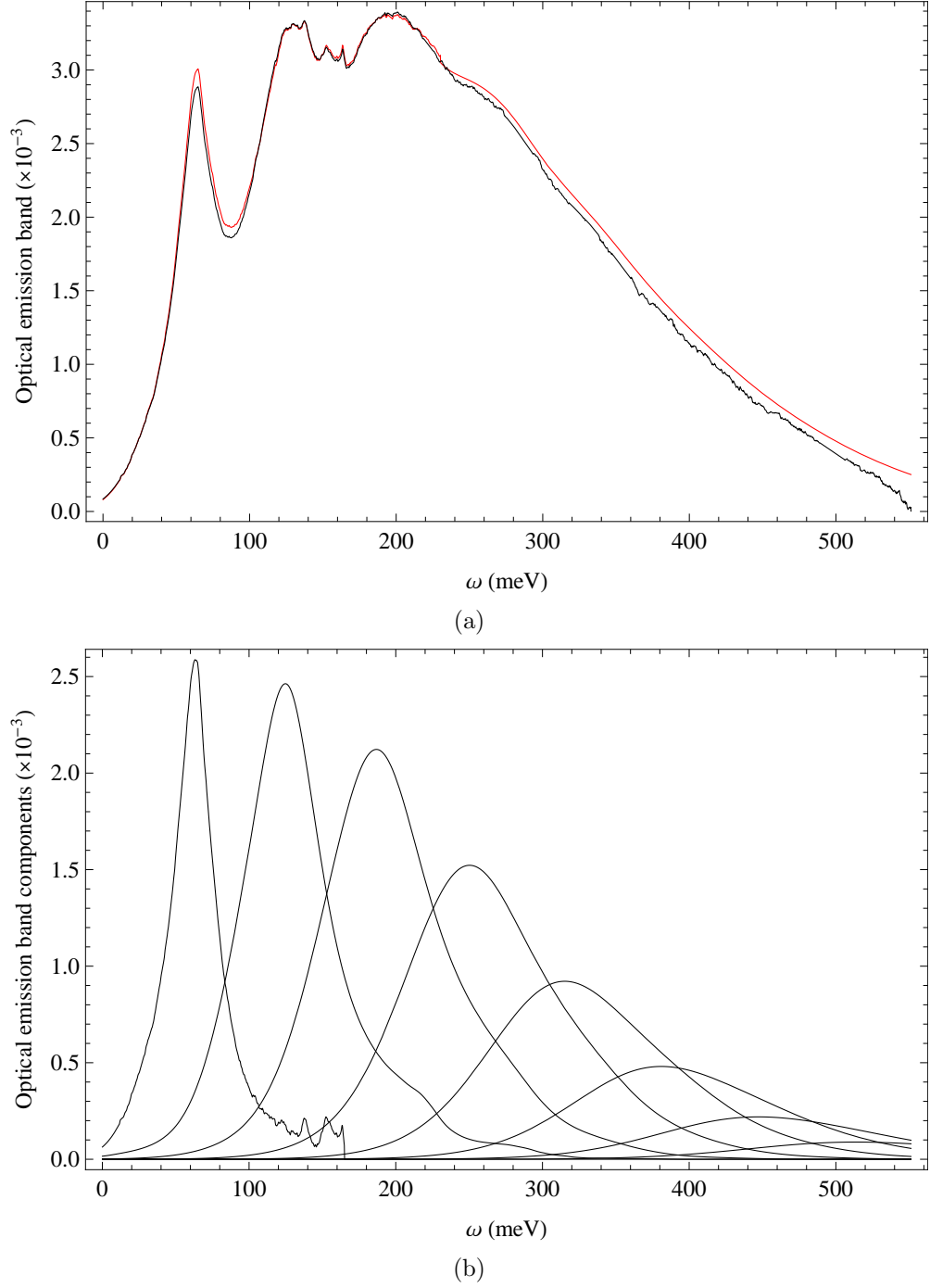

(a)

(b)

FIG. S5. (a) The normalized observed optical emission PSB (black) and the calculated PSB (red) obtained using the optical emission $g(\omega)$ and (6). (b) The calculated $n$-phonon components (increasing from $n$=1 from left to right) of the optical emission PSB. The sum of the components equals the calculated PSB depicted in (a).

phonon modes of perfect diamond, we can explain the features of the optical $g(\omega)$ via a comparison with the diamond phonon dispersion curves along high symmetry directions ([111], [100] and [110]) and the phonon density of states (DOS). As the NV$^-$ electronic charge is centered on the nearest-neighbor nuclei of the vacancy (NNV), we expect that the NV$^-$ center will interact strongly with modes where the NNV undergo the largest relative displacement, which occurs when the NNV vibrate directly out of phase. As the NNV occupy equivalent Bravais lattice sites, this phase condition can be expressed as $\vec{k} \cdot \vec{R} = \pi$, where $\vec{k}$ is the phonon wavevector and $\vec{R}$ is the lattice vector connecting two NNV. For a given phonon wavevector direction, the phase condition is satisfied by a single wavevector magnitude. The magnitudes corresponding to the [111], [100] and [110] directions are $k = \frac{a}{R}$, $\frac{a}{R}$ and $\frac{2a}{3R}$, respectively, where $a$ is the distance between equivalent Bravais lattice sites in the perfect diamond lattice and $R$ is the distance between two NNV.

As expected, Fig. S6a shows that features of the optical $g(\omega)$ approximately correspond to features in the DOS and to points along the phonon dispersion curves at which the wavevector in the [111] direction has a particular magnitude. Fig. S6 also depicts similar comparisons with the [100] and [110] dispersion curves. By considering all of the comparisons together, it is clear that one can attribute all of the optical $g(\omega)$ features to points along the phonon dispersion curves whose wavevector magnitudes are related by the parameter $R/a \sim 1.2$. This semi-empirical result is in reasonable agreement with the ratios $R/a \sim 1.1 - 1.3$ obtained in previous *ab initio* calculations [14, 15]. Hence, the elements of the $^1A_1$ and $^3A_2$ vibronic structures observed in the IR absorption and optical emission PSBs can be directly related to specific phonon subgroups of diamond.

———————

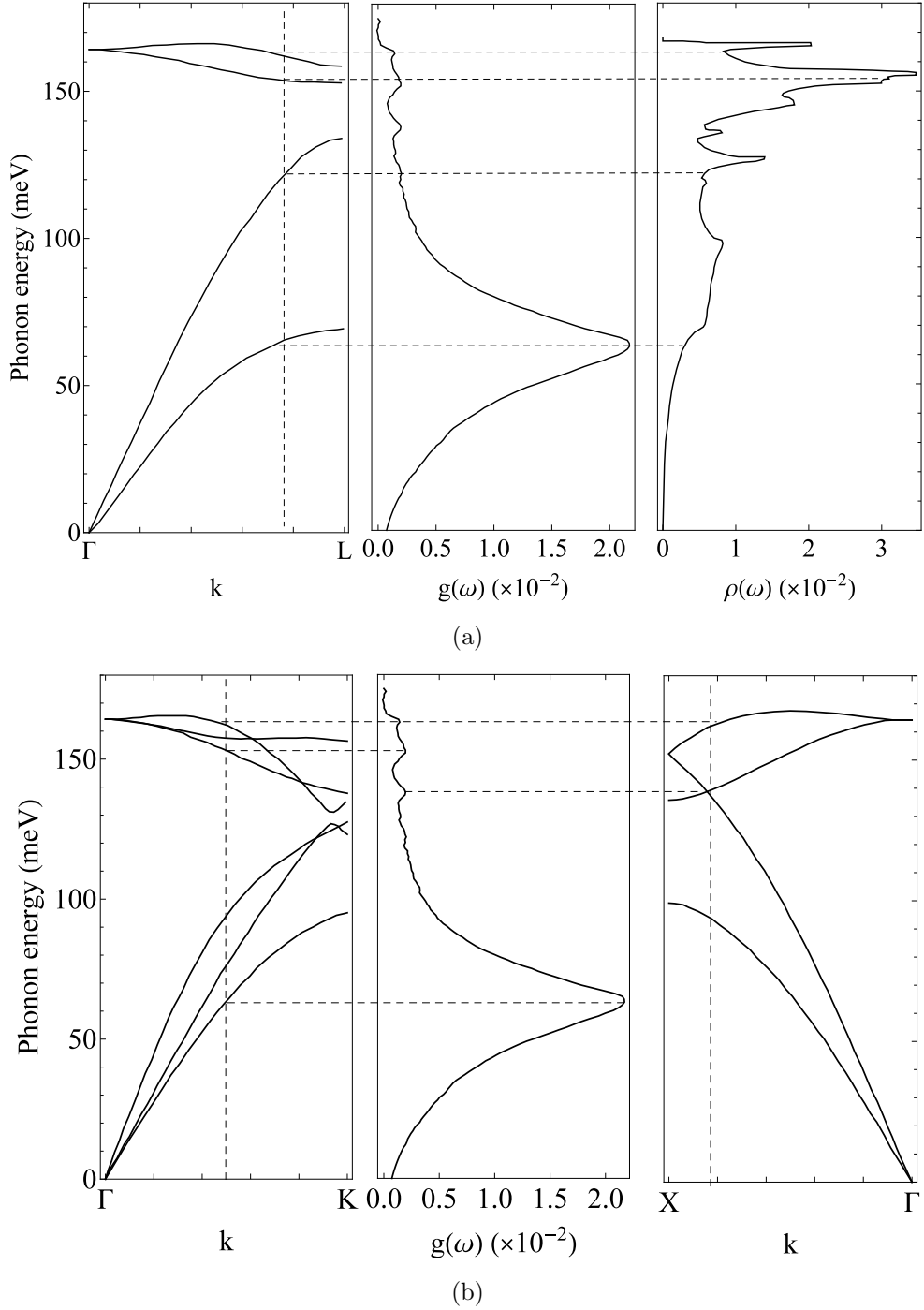

FIG. S6. (a) The comparison of the optical emission $g(\omega)$ with the perfect diamond phonon DOS $\rho(\omega)$ [16] and dispersion curves of phonons with wavevectors in the [111] direction [16]. (b) Comparison with the dispersion curves of phonons with wavevectors in the [110] direction (left) and [100] direction (right) [16]. Horizontal dashed lines connect related features in adjacent plots and the vertical dashed line connects points with the same wavevector magnitude.

TABLE S2. The vibrational parameters of the NV$^-$ center's electronic levels as derived using the current model of the center. The parameters are expressed in terms of the Coulomb coupling parameter $\kappa_E$ (refer to [10] for further details), $\kappa_A$, and molecular-orbital reduced matrix elements of the linear and quadratic electron-phonon interactions. The explicit expressions for the reduced matrix elements are contained in Table S3.

| | $^3A_2$ | $^1E_x$ | $^1E_y$ | $^1A_1$ | $^3E_x$ | $^3E_y$ |
|---|---|---|---|---|---|---|
| $a_{n,\alpha,A_1}$ | 0 | $\kappa_E^2 a_{\alpha,A_1}$ | $\kappa_E^2 a_{\alpha,A_1}$ | $2\kappa_A^2 a_{\alpha,A_1}$ | $a_{\alpha,A_1}$ | $a_{\alpha,A_1}$ |
| $a_{n,\alpha,A_2}$ | 0 | 0 | 0 | 0 | 0 | 0 |
| $a_{n,\alpha,E,x}$ | 0 | $2\kappa_E a_{\alpha,2,E}$ | $-2\kappa_E a_{\alpha,2,E}$ | 0 | $a_{\alpha,1,E}$ | $-a_{\alpha,1,E}$ |
| $a_{n,\alpha,E,y}$ | 0 | 0 | 0 | 0 | 0 | 0 |
| $b_{n,\alpha,A_1}$ | 0 | $\kappa_E^2 b_{\alpha,A_1}$ | $\kappa_E^2 b_{\alpha,A_1}$ | $2\kappa_A^2 b_{\alpha,A_1}$ | $b_{\alpha,A_1}$ | $b_{\alpha,A_1}$ |
| $b_{n,\alpha,A_2}$ | 0 | $\kappa_E^2 b_{\alpha,A_2}$ | $\kappa_E^2 b_{\alpha,A_2}$ | $2\kappa_A^2 b_{\alpha,A_2}$ | $b_{\alpha,A_2}$ | $b_{\alpha,A_2}$ |
| $b_{n,\alpha,E,x}$ | 0 | $\kappa_E^2 b_{\alpha,E}$ | $\kappa_E^2 b_{\alpha,E}$ | $2\kappa_A^2 b_{\alpha,E}$ | $b_{\alpha,E}$ | $b_{\alpha,E}$ |
| $b_{n,\alpha,E,y}$ | 0 | $\kappa_E^2 b_{\alpha,E}$ | $\kappa_E^2 b_{\alpha,E}$ | $2\kappa_A^2 b_{\alpha,E}$ | $b_{\alpha,E}$ | $b_{\alpha,E}$ |

TABLE S3. The reduced matrix elements of the linear and quadratic electron-phonon interactions expressed in terms of the NV$^-$ center's molecular orbitals ($a_1$, $e_x$ and $e_y$). Refer to [10] for further details of the molecular orbitals and calculation of reduced matrix elements.

$$a_{\alpha,A_1} = \langle e| \left.\frac{\partial \hat{H}_e}{\partial Q_{\alpha,A_1}}\right|_0 |e\rangle - \langle a_1| \left.\frac{\partial \hat{H}_e}{\partial Q_{\alpha,A_1}}\right|_0 |a_1\rangle$$

$$a_{\alpha,1,E} = \langle e| \left.\frac{\partial \hat{H}_e}{\partial Q_{\alpha,E}}\right|_0 |e\rangle$$

$$a_{\alpha,2,E} = \langle a_1| \left.\frac{\partial \hat{H}_e}{\partial Q_{\alpha,E}}\right|_0 |e\rangle$$

$$b_{\alpha,A_1} = \langle e| \left.\frac{\partial^2 \hat{H}_e}{\partial Q_{\alpha,A_1}^2}\right|_0 |e\rangle - \langle a_1| \left.\frac{\partial^2 \hat{H}_e}{\partial Q_{\alpha,A_1}^2}\right|_0 |a_1\rangle$$

$$b_{\alpha,A_2} = \langle e| \left.\frac{\partial^2 \hat{H}_e}{\partial Q_{\alpha,A_2}^2}\right|_0 |e\rangle - \langle a_1| \left.\frac{\partial^2 \hat{H}_e}{\partial Q_{\alpha,A_2}^2}\right|_0 |a_1\rangle$$

$$2b_{\alpha,E} = \langle e| \left.\frac{\partial^2 \hat{H}_e}{\partial Q_{\alpha,E,x}^2} + \frac{\partial^2 \hat{H}_e}{\partial Q_{\alpha,E,y}^2}\right|_0 |e\rangle - \langle a_1| \left.\frac{\partial^2 \hat{H}_e}{\partial Q_{\alpha,E,x}^2}\right|_0 + \frac{\partial^2 \hat{H}_e}{\partial Q_{\alpha,E,y|0}^2} |a_1\rangle$$


\end{document}